\documentclass[12pt]{iopart}
\usepackage{iopams}
\usepackage[portuges,english]{babel}
\usepackage[T1]{fontenc}
\usepackage{color}
\usepackage{amssymb,amsthm,latexsym}
\usepackage{graphics,graphicx,epsfig}
\usepackage{multicol}
  \usepackage{subfigure}

\begin{document}

\title[Calibration methods for spatial Data]{Calibration methods for spatial Data }
\author{M A Amaral Turkman$^1$, K F Turkman$^1$, P de Zea Bermudez$^1$, S Pereira$^1$, P Pereira$^2$ and  M Carvalho$^3$ }

\address{$^1$ Centro de Estat\'{\i}stica e Aplica\c{c}\~{o}es,
Faculdade de Ci\^{e}ncias, Universidade de Lisboa}
\address{$^2$ Instituto Polit\'{e}cnico Set\'{u}bal and Centro de Estat\'{\i}stica e Aplica\c{c}\~{o}es,
 Universidade de Lisboa}
\address{$^3$ University of Edinburgh and Centro de Estat\'{\i}stica e Aplica\c{c}\~{o}es,
 Universidade de Lisboa}
\ead{maturkman@fc.ul.pt}

\begin{abstract}
In an environmental framework, extreme values of certain spatio-temporal processes, for example wind speeds, are the main cause of severe damage in property, such as electrical networks, transport and agricultural infrastructures. Therefore, availability of accurate data on such processes is highly important in risk analysis, and in particular in producing probability maps showing the spatial distribution of damage risks.  Typically, as is the case of wind speeds, data are available at few stations with many missing observations and consequently simulated data are often used to augment information, due to simulated environmental data being available at high spatial and temporal resolutions. However, simulated data often mismatch observed data, particularly at tails, therefore calibrating and bringing it in line with observed data may offer practitioners more reliable and richer data sources. Although the calibration methods that we describe in this manuscript may equally apply to other environmental variables, we describe the methods specifically with reference to wind data and its consequences. Since most damages are caused by extreme winds, it is particularly important to calibrate the right tail of simulated data based on observations. Response relationships between the extremes of simulated and observed data are by nature highly non-linear and non-Gaussian, therefore data fusion techniques available for spatial data may not be adequate for this purpose. Although, our ultimate goal is the development of statistical methods for data fusion and calibration that can extrapolate beyond the range of observed data---into the tails of a distribution---in this work we will concentrate on calibration methods for the whole range of data. After giving a brief description of standard calibration and data fusion methods to update simulated data based on the observed data, we propose and describe in detail a specific conditional quantile matching calibration method and show how our wind speed data can be calibrated using this method. We also briefly explain how calibration can be extended specifically to data coming from the tails of simulated and observed data, using asymptotic models and methods suggested by extreme value theory
\end{abstract}
\maketitle

\section{Introduction}
Extreme values of certain spatio-temporal processes, such as wind speeds, are the main cause of severe damage in property, from electricity distribution grid to transport  and agricultural infrastructures. Accurate assessment of causal relationships between environmental processes and their effects on risk indicators, are highly important in risk analysis, which in return depends on sound inferential methods as well as on good quality informative data. Often, information on the relevant environmental processes comes from monitoring networks, as well as from numerical-physical models (simulators) that typically solve a large set of partial differential equations, capturing the essence of the physical process under study (Skamarock \emph{et al.} 2008, Cardoso \emph{et al.} 2013).  In general, monitoring networks are formed by a sparse set of stations, whose instrumentation are vulnerable to disruptions, resulting in data sets with many missing observations, whereas, simulated data from numerical simulators typically supply average yield of the process in grid cells of pre-specified dimensions, often at high resolutions, spanning large spatial domains, with no missing observations. However, simulated data typically mismatch and misaligned observed data, therefore calibrating it and bringing it in line with observed data may supply modellers with more reliable and richer sources of data. Data assimilation methods, namely combining data from multiple sources, are well known in environmental studies, with data often being used to generate initial boundary conditions for the numerical  simulators (Kalnay, 2003).  There is a very rich statistical literature on data assimilation and data fusion with the objective of enriching the information for inference (Fuentes and Raftery 2005, Berrocal \emph{et al.} 2012, Zidek \emph{et al.} 2012, Berrocal \emph{et al.} 2014, McMillan \emph{et al.} 2010). These statistical methods are often based on Bayesian hierarchical methods for space time data (Banerjee \emph{et al.} 2004) and are constructed around the idea of relating the monitoring station data and the simulated data using spatial linear models with spatially varying coefficients (Berrocal 2019). Since these relations involve data measured at different spatial resolutions, the models often are called downscaler models (Berrocal \emph{et al.} 2012). The principal objective of these downscaler models is to relate observations measured at different space resolutions using spatial linear models. However, as a by-product, these models can be used for calibrating one set of data as a function of the other, as it will be explained later.

The motivation behind this present work has its roots in a
consulting work done for  a major electricity producer and distributor.
  The electricity grid constantly faces disruptions due to damages
in the distribution system, with heavy economic losses. These damages and consequent disruptions occur due to a  combination of many
factors such as topography and  precipitation, however extreme winds and storms are the main cause of such damages.  Risk maps that indicate likely places of costly disruptions in electric grids are important decision support tools for administering the power grid and are particularly useful in deciding if   costly corrective actions should  be taken to improve
structures. It is natural that these risk maps should be based  primarily  on observed wind speeds among other factors and it was decided that daily maximum wind speeds should be used as proxy information.  Hence, such risk maps can be interpreted as vulnerability  maps of electricity grid to extreme wind speeds, expressed in terms of probability.  However generating such  maps depends on reliable wind data at fairly high spatial and temporal resolutions.

     The data available for this particular study corresponds to simulated wind speeds from a simulator (The WRF model, version 3.1.1) at a regular grid of 81ksq grid cell size, obtained at 10 minutes interval from 2006-2013; however only daily maximum wind speed will be used.  Observed daily maximum wind speed is also available during the same period of time, from 117 stations in Portugal mainland, but missing observations reach to 90\% in some stations. Only around one third of the stations have less than 30\% missing observations. There is an additional challenge: although simulated and observed data are similar in the bulk
of the distribution,  they quite often mismatch  at extreme values. Therefore,
 adequate methods of data
fusion  and calibration, can be used to combine these two different sources of data and may   provide information which is more reliable from a spatial point of
view and  produce more accurate probability maps showing the spatial
distribution of damage risks.
Since electricity grid damages are ultimately caused by extreme wind speeds, ultimate aim should be  to develop  statistical methods for data fusion and calibration  that can extrapolate beyond the
range of observed data  into the  tails of a distribution.  However, in this manuscript,  we make a review of statistical fusion and calibration  methods  for the whole range of data. Calibration methods that extend beyond the range of data will be reported elsewhere.
Our objective is to explore several methods to model the relationship between simulated and observed wind speeds at observation sites, so as to extrapolate this relationship in space at grid cell or county level resolution. In other words, more than imputing missing observations, we want to use simulated wind speeds for risk assessment, after being calibrated, i.e., brought in line with observed wind speeds.
    After giving a brief description of standard calibration and data fusion methods to update simulated data based on the observed data, we will propose and describe in detail a specific conditional quantile matching calibration method  and show how our wind speed data can be calibrated using this method.  We also briefly explain how calibration can be extended specifically to data coming from the tails of simulated and observed data, using asymptotic models and methods suggested by extreme value theory.

The outline of the paper is as follows: In section \ref{cali}, we give an
    overview of statistical calibration methods. In section \ref{NAV}, we report a new approach
 for calibration through a conditional quantile matching calibration method (Pereira \emph{et al}., 2019), using an extended Generalized Pareto distribution (Naveau \emph{et al.}, 2016) for the simulated and observed data, adequate for calibrating simultaneously  the bulk and the tails of the distribution.
Finally, in section \ref{wind},
 this method will be exemplified using a wind speed data. Further discussion and conclusions are in section \ref{disc}.


\section{Statistical Calibration methods; an overview} \label{cali}

Calibration plays a crucial role in almost all areas of experimental sciences and can be defined in a nutshell as  “the comparison between measurements - one of known magnitude or correctness made or set with one device and another measurement made in as similar a way as possible with a second device “ (Wikipedia).  When measurements are obtained under  random environments, then statistical methods and models have to be employed to compare such sets of measurements. In the simplest case of random experiments involving linear calibration,  linear regression models are used as models. The univariate calibration is then defined as inverse regression problem (Lavagnini and Magno, 2007). These models and consequent calibration techniques  are, inevitably, restricted to  uncorrelated repeated observations or dependent  Gaussian structures, simplifying immensely the problem (Aitchison and Dunsmore, 1976). However, more often than not, even in designed experiments, the response relationships are mostly nonlinear and therefore  Gaussian structures are hardly justifiable as models.  Nonlinear calibration then typically have to be formulated by conditional specification of distributions, and consequently  substantial amount of numerical integration and approximations are needed.

Little is known on calibration of environmental data sets displaying nonlinear, non-Gaussian structures in a spatio-temporal setting. In these cases, defining calibration through spatial linear models as inverse regression problems will oversimplify the structures and will not be adequate. Often calibrating simulated data based on observed data is  done by calibration of the simulator, namely the numerical-physical model, using Bayesian methods (Kennedy and O’Hagan, 2001,  and Wilkinson, 2010). These generic methods are based on the following general ideas: The simulator first is approximated by a linear parametric  emulator, and using Bayesian arguments,  data are used to convert prior knowledge on these model parameters to posterior distributions. The newly generated data from this approximate emulator  is  then treated as the calibrated data.

There are many statistical calibration methods for different purposes and based on different paradigms.  We can classify these methods into
\begin{enumerate}
\item[(i)] Quantile matching-based approaches,
\item[(ii)] Inverse regression,
\item[(iii)]  Simulator--emulator-based approaches,
\item[(iv)]  Data assimilation/data fusion.
\end{enumerate}

Before describing these methods, we give here some basic notation.

 We denote by  $Y(s,t)$ and $X(s,t)$,  respectively the observed and simulated  wind speeds at location $s\in \mathbb{R}^2$ and time $t$.
To simplify notation, often  we will use $Y$ and $X$ for observed and simulated wind speeds when data are used without any space-time  reference. Typically  $X$ are  simulated over a regular grid, say  $B$,  often  represented by  points $s_B$ which correspond to  the centroid of the grid cells, whereas   $Y$ are observed in stations located at different spatial points $s$.

\subsection{Quantile matching-based approaches}

 For the time being, if we  ignore totally space-time variations and dependence structures, calibration can be seen as a
simple scaling making use of marginal distributions fitted respectively to $X$ and $Y$ (CDF transform method, Michelangeli \emph{et al}, 2009).

Suppose we have a set of $n$ observed $y_i$ and simulated $x_i$ , $i=1,...,n$ data. Let $F_Y$ and $F_X$ be, respectively, the distribution functions of $Y$ and $X$. Then the new calibrated (scaled) data $x_i^*$ is defined as
 \begin{equation}
 x_i^*=F^{-1}_Y(F_X(x_i)), \quad i= 1, \ldots,n. \label{qmb}
\end{equation}

\begin{figure}
  \centering
  \includegraphics[width=10cm]{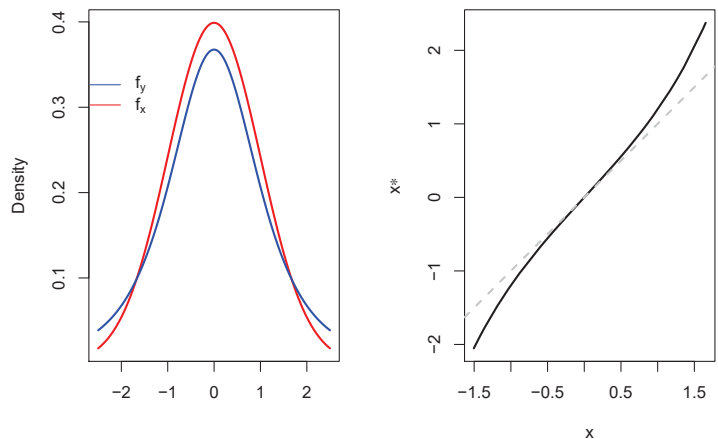}
  \caption{Illustration of the quantile matching approach}\label{fig1}
\end{figure}

 Since
$$P(X^*\le z)=F_Y(z),$$
calibrated data has the same distribution as the observed data. Note that if $F_X=F_Y$ then $x_i^*=x_i$. Figure \ref{fig1} depicts the result of applying this calibration method  when $Y$ follows a Student distribution with 3 degrees of freedom, and $X$ follows a standard normal distribution.

 This calibration method depends on the marginal distributions of the random variables involved $Y$ and $X$ and hence it does not
make use of the expected strong dependence between the two sets of data.

An ideal calibration should involve the joint distribution of $Y$ and $X$ defined in some way. A possibility is the use of  a conditional quantile matching approach,  which will be described in section \ref{NAV}.  Further, in the same  section,   we also introduce
an extension  to  cover space-time non-homogeneity  by scaling (calibrating) the data from
\begin{equation}
 x^*(s,t)=F^{-1}_{Y(s,t)}(F_{X(s,t)}(x(s,t)),\label{eqmb}
\end{equation}
assuming  marginal distributions of $Y(s,t)$ and $X(s,t)$  for every $s$ and $t$.

 These distributions will be estimated by fitting them and considering the
 parameters  as smooth functions of spatially and temporarily varying covariates and space-time latent processes as in section \ref{wind}.

    \subsection{Inverse regression}

 Calibration is usually seen as a method of adjusting the scale of a measurement instrument on the basis of an informative experiment. As such, it is seen as an inverse regression problem. However, there are several problems associated with this approach (see, e.g. Kang \emph{et al.}, 2017).
 Aitchison and Dunsmore (1975) approach the problem from a Bayesian perspective by defining the \textit{calibrative distribution}. See also (e.g. Racine-Poon, 1988, Osborne, 1991 and Muehleisen and Bergerson, 2016).

According to Aitchison and Dunsmore's proposal, a parametric model is fitted to a random vector $(X,Y)$, where, e.g. $X$ is the random variable representing a  measurement obtained in a laboratory and $Y$ the random variable representing the measurement  obtained in a field experiment. This parametric model is defined through a conditional specification such that
$$f_{(X,Y)}(x,y\mid \psi,\theta)=f_{X\mid Y}(x\mid y,\psi)f_Y(y\mid \theta).$$ There are two parameters involved, namely  the arrival parameter $\psi$ and the structural parameter $\theta$. A further assumption is that the initial sources of information are stochastically independent, so that
$$p(\psi,\theta)=p(\psi)p(\theta).$$

 Now suppose that one has a future laboratory experiment resulting in a value $x_f$  and  that further experiments follow the same pattern of arrival as the original trials $(\mathbf{x},\mathbf{y})=(x_i,y_i, i=1,...,n)$. Now the data available is $(x_i,y_i, x_f), i=1,...,n$ and the unknowns are $(y_f,\psi,\theta)$. The objective is to obtain the predictive distribution (called in this case the \emph{calibrative distribution}) for the corresponding field experiment $Y_f$, which is simply obtained by integrating out $\psi$ and $\theta$,

$$p(y_f\mid \mathbf{x},\mathbf{y},x_f)=\int f_X(x_f\mid y_f,\psi)p(y_f,\psi,\theta\mid x_f,\mathbf{x},\mathbf{y})d\psi d\theta  $$
 where
 $$p(y_f,\psi,\theta\mid x_f,\mathbf{x},\mathbf{y}) \propto f_X(x_f\mid y_f,\psi) f_Y(y_t\mid \theta)\prod_{i=1}^{n} f_X(x_i\mid y_i,\psi)f_Y(y_i\mid \theta)p(\psi)p(\theta),$$
 assuming that the trial records are independent.

Generalizing to the  situation under study, without considering space and time dependence,
for a model $[X\mid Y][Y]$, the objective is to obtain the distribution $$Y(s_0)\mid x(s_0),x(s^*),y(s^*)$$ for an unknown $Y(s_0)$ based on the observed and simulated data $(x(s^*),y(s^*))$ on $N$ stations and the simulated value $x(s_0)$.

    \subsection{Simulator--emulator-based approaches}

 Kennedy and O'Hagan (2001)  describe calibration  as statistical postprocessing of simulator deterministic forecast.
  They assume a computer model describing some physical system, that is a function of variable inputs $x$ that can be measured and calibration inputs $\nu$ needed to run the model, but whose values are not
known in the experiment. The output of the computer model is then assumed to be some function of the inputs, say $\eta(x,u)$. Observations from the field experiment are assumed to have been observed at $u = \theta$ and, possibly at different values of $x$. The model for observations for known input variables $x$ is then assumed to be a function of the computer model output $\eta(x,\nu)$, of a
true underlying process $\xi(x)$  and some  model inadequacy described by $\delta(x)$.
The objective is to estimate the calibration settings $\theta$ consistent with the field experiments and the computer model.

Sigrist \emph{et al.} (2015) give detailed description of  stochastic versions of space-time advection-diffusion PDE's and their solutions as models for emulators and describe a method of postprocessing simulated data.

However  simulator--emulator-based approaches assume  detailed information of how emulators work in terms of a set of parameters, which is not the case in most situations related to climate models.

 \subsection{Data fusion and calibration}

  Data assimilation or data fusion methods are used to combine different sources of information in order to obtain more accurate results. A recent review on data assimilation is given in  Berrocal (2019) together with many references.

  Among these methods, the interest lies in statistical approaches to data assimilation/data fusion and
  in particular to  hierarchical Bayesian models (HBM)  for combining monitoring data and computer model output.

  There are basically two different approaches regarding these methods. The Bayesian melding proposed by Fuentes and Raftery (2005) assumes that there is a true latent point-level process $Z(s)$ (\emph{a GP with spatially varying mean and non-stationary covariance function}) to which both the observed $Y(s)$ and simulated $X(s)$ processes are linked to.
 The observed values are governed by this latent process  with a random error and simulated values are  expressed as a linearly calibrated integral over a grid cell (scaled by the area of the cell) of the latent point-level process,
 $$Y(s)=Z(s)+\epsilon(s) \qquad X(s)=a(s) + b(s)Z(s) + \delta(s),$$
where $\delta(s)$  explains the random deviation
at location $s$ with respect to the underlying true process $Z(s)$.
 The aim is to obtain the posterior
predictive distribution of the truth $Z$ at a new site $s^*$. However, the misalignment of the two processes involved brings computational difficulties in the implementation. Foley and Fuentes (2008) apply this type of modelling to hurricane surface wind prediction and
McMillan \emph{et al.} (2010) propose a spatio-temporal extension of this Bayesian melding approach.

  The other approach suggested by  Berrocal \emph{et al.} (2010) is a Bayesian hierarchical downscaler model.  They consider a spatial linear model relating the monitoring station data and the computer model output, with spatially varying coefficients which are in turn modeled as Gaussian processes (GPs). These models offer the advantage of local calibration of the numerical model output without incurring in problems due to the dimensionality of the computer model output, since they are only fitted at the  grid cells where the monitoring stations reside. An extension to this downscaler model, by borrowing information from neighboring grid cells, was introduced by Berrocal \emph{et al.} (2012).
  The proposed approach, the downscaler model   for the observations and simulated data from the computer model is

$$Y(s) =\beta_0+\beta_0(s)+\beta_1+\beta_1(s)x(B)+e(s),\quad e(s)\quad i.i.d.\quad N(0, \tau^2),$$ with   $B$ the grid cell containing $s$
 $$X(B)= \mu + V(B) +\eta(B), \quad \eta(B) \quad i.i.d. \quad N(0, \sigma^2)$$ where $V(B)$ is a GP model with a ICAR structure (Rue and Held, 2005), and $\beta_0(s)$ is modeled as a mean-zero GP with exponential covariance structure.

A smoothed version is possible considering
$$Y(s) =\beta_0+\beta_0(s)+\beta_1(\mu+V(B))+e(s),\quad e(s)\quad i.i.d. \quad N(0, \tau^2)$$ with   $B$ the grid cell containing $s$.

 The aim is to obtain the predictive distribution of $Y$ and its expected value at grid cell level.

They also considered a smoothed downscaler using spatially varying random weights and a
space-time extension.

\section{Calibration methods for bulk and tails}\label{NAV}

Pereira \emph{et al.} (2019) develop  a covariate-adjusted version of the  quantile matching-based approach as in (\ref{qmb}) where the distributions of simulated and real data change along a covariate. At the same time they suggest a regression method
that simultaneously models the bulk and the (right) tail of  the distributions involved
 using the extended Generalized Pareto distribution  (EGPD) (Naveau \emph{et al.}, 2016) as a model for both the simulated and observed data.

 Under fairly general conditions, according to the asymptotic theory of extremes, the generalized Pareto distribution (GPD) appears as a natural model for the right tail of a distribution, by focusing on the  excesses over a high but fixed threshold. Here, the choice of this threshold plays a very important role in inference, ignoring the part of the data that lie below this threshold. See, for example, Beirland \emph{et al.} (2004). The EGPD modelling strategy suggested by Naveau \emph{et al} (2016) avoids this selection problem, as we will see in next section.

 In what follows we propose an extension of this conditional quantile matching calibration for the bulk and tails, to spatial temporal data.

\subsection{ Naveau \emph{et al} (2016) EGPD models}
Naveau \textit{et al.}~(2016) suggest an extension of Generalized Pareto  model tailored for both the bulk and tails, and---contrarily to most methods for extremes--- does not require a threshold to be selected.  The objective of this extension is to generate a new class of distributions with GPD type tails consistent with extreme value theory, but also flexible enough to model efficiently the main bulk of the observed data without complicated threshold selection procedures.

Let $Y$ be a positive random variable with cumulative distribution function defined as:
 \begin{equation*}
  F_Y(y \mid \theta) = G\bigg(H(y\mid\xi,\sigma )\bigg),
  \label{qr}
\end{equation*}
where $G$ is a  function obeying some general assumptions (see Naveau \emph{et al.}, 2016) and $H$ is the cumulative distribution function of a Generalized Pareto distribution (GPD), that is
\begin{equation*}
  H(y\mid\xi,\sigma ) =
\left\{
  \begin{array}{ll}
     1 - (1 + \frac{\xi}{\sigma} y)_{+}^{-1/\xi}, & \hbox{$\xi  \neq 0$.} \\
    1 - \exp(-\frac{y}{\sigma}), & \hbox{$\xi = 0$ .}
  \end{array}
\right.
\end{equation*}
 with $\sigma>0$, and $y>0$ if $\xi\geq 0$ and $y<-\frac{\sigma}{\xi}$ if $\xi<0$. The parameter $\sigma$ is a dispersion parameter while $\xi$ is a shape parameter controlling the rate of decay of the right tail of a distribution (de Zea Bermudez and Kotz, 2010).

  Naveau \emph{et al.} (2016)  consider four  forms of $G(u)$ resulting in four different classes of distributions.

\subsection{Spatio temporal conditional quantile matching calibration for the bulk and tails}
Here we  use one of the forms, namely, $G(u)=u^\kappa$ where $\kappa$ is a parameter controlling the shape of the lower tail, although the theory can be easily extended to any of the other forms of the $G$ function.

Let us assume that both random variables  $X$  and  $Y$  are space-time dependent  and we want to calibrate $X$ based on $Y$. As in (\ref{eqmb}) the calibrated data is given as
\begin{equation*}
 x^*(s,t)=F^{-1}_{Y(s,t)}(F_{X(s,t)}(x(s,t)),
\end{equation*}
Now assume further that both random variables are distributed as an EGPD with different parameters. In order to better accommodate for the situation $\xi<0$ we make a transformation $\delta=-\frac{\sigma}{\xi}$. Hence, for $\xi_x\neq 0$
\begin{equation}
F_{X(s,t)}(x(s,t)\mid \delta_x(s,t), \xi,\kappa_x)=
   \left(1 - \left(1 - \frac{1}{\delta_x(s,t)} x(s,t)\right)_{+}^{-1/\xi_x}\right)^{\kappa_x},  \label{fx}
\end{equation}
and assuming as well $\xi_y\neq 0$
\begin{equation}
F_{Y(s,t)}(y(s,t)\mid \delta_y(s,t), \xi_y,\kappa_y)=
   \left(1 - \left(1 - \frac{1}{\delta_y(s,t)} y(s,t)\right)_{+}^{-1/\xi_y}\right)^{\kappa_y}. \label{fy}
\end{equation}

Although it is assumed that these random variables are conditional independent, a dependence structure is introduced through the transformed space-time dependent parameters $\delta_x, \delta_y$  by modelling them as a function of a common random spatio-temporal process, in a Bayesian hierarchical modelling framework.

As an exemplification of this modelling strategy, in the next section, we will built a Bayesian hierarchical model for the wind speed data.

\section{Bayesian hierarchical model for  the wind speed data}\label{wind1}
 A preliminary data analysis of the wind speed data used in this study, shows that observed and simulated data are consistent with the case $\xi<0$ and hence, the distributions for $X$ and $Y$ will have an end-point characterized by the respective  parameter $\delta$.

We assume that the observed data $\{Y(s_i,t_j), i=1,...,N; j=1,...,T\}$, with $N$ the number of stations with observed data in the  study period and $T$  the length of the time period,  follow a distribution as in (\ref{fy}), where
  $\delta_y(i,j)\sim Exp(\lambda_y(i,j)) ,$ $\delta_y(i,j)>\max(y),$ i.e., follows a shifted exponential distribution with
  $\log(\lambda_y(i,j))=\beta_y+ W(s_i)+Z(t_j),$ and $W$ follows a  Multivariate Gaussian process, defined on the space,  $W\sim MVN(0,\tau_W\Sigma_W)$.
The  matrix $\Sigma_W$ has diagonal elements equal to 1 and off-diagonal elements, $\Sigma_{i\ell}=f(d_{i\ell};\alpha)$, where $f(.;.)$ is a function of  $d_{i\ell}$,  the centroids' distance of every two stations $s_i$ and $s_\ell$, and  $\alpha$ a parameter representing the radius of the 'disc' centred at each  $s$. The  $\tau_W$ is a precision parameter. For the temporal random process we assume a  random walk process of order 1, $Z\sim MVN(0,\tau_Z\Sigma_Z)$,  where $\tau_Z$ is a precision parameter and $\Sigma_Z$ is a matrix with a structure reflecting the fact that any two increments $z_i-z_{i-1}$ are independent (Rue and Held, 2005).

We assume, as well, that the simulated data $\{X(s_i,t_j), i=1,...,N_s; j=1,...,T\}$ follow a distribution as in (\ref{fx})
 with $N_s$ total number of stations, where the model for $\delta$  shares the same latent processes $W$ and $Z$ with the model for the observed data, such that $\delta_x(i,j)\sim Exp(\lambda_x(i,j)),$ $\delta_x(i,j)>\max(x),$  with
 $\log(\lambda_x(ij))=\beta_x+ W(s_i)+Z(t_j).$

To complete the Bayesian hierarchical model we consider the following prior specification for the parameters and hyperparameters of the models

  $\beta_y,\beta_x$ i.i.d. $N(0,0.01)$,

  $\kappa_y,\kappa_x$ i.i.d. $Ga(0.05,0.05)$,

  $\xi_y, \xi_x$ i.i.d. $U(-0.5,0)$,

  $\tau_W, \tau_Z$ i.i.d. $Ga(1,0.1)$,

  $\alpha \sim U(0.1,0.5)$.

Finally, the calibrated values are obtained as the
mean of the predictive distribution of $F^{-1}_Y(F_X(x(s_i,t_j))$
 at $s_i, i=1,...,N_s$ and time $t_j,j=1,...,T$.

\section {Application to wind speed data }\label{wind}

\begin{figure}[!ht]\centering
\includegraphics[width=10cm]{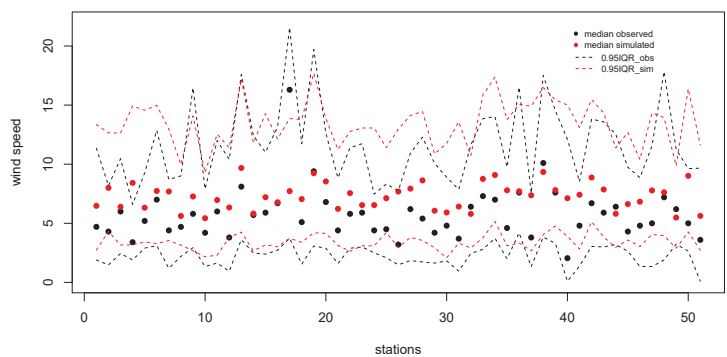}
\caption{\label{turkman:fig1} Median of observed  and simulated wind speeds for the 51 stations, and the 95\% IQR wind speeds by station (dashed lines).}
\end{figure}

We used observed and simulated wind speed data from the period 01/01/2013 to 28/02/2013, so $T=59$. There are $N=51$ stations where we have both observed and simulated daily maximum wind speeds. Additionally we have extra 66 stations with simulated values for the maximum daily wind speeds, so that $N_s=117$. In Figure \ref{turkman:fig1} we depict the median of observed  and simulated wind speeds for the 51 stations together with the
2.5\% and 97.5\% empirical quantiles (95\% IQR).

The model was implemented in \texttt{R2OpenBUGS} (Sturtz \emph{et al.} (2005). In Table \ref{tab1} we show the summary statistics for the  marginal posterior distributions of the parameters of the model.

\begin{figure}[!ht]\centering
\includegraphics[width=7cm]{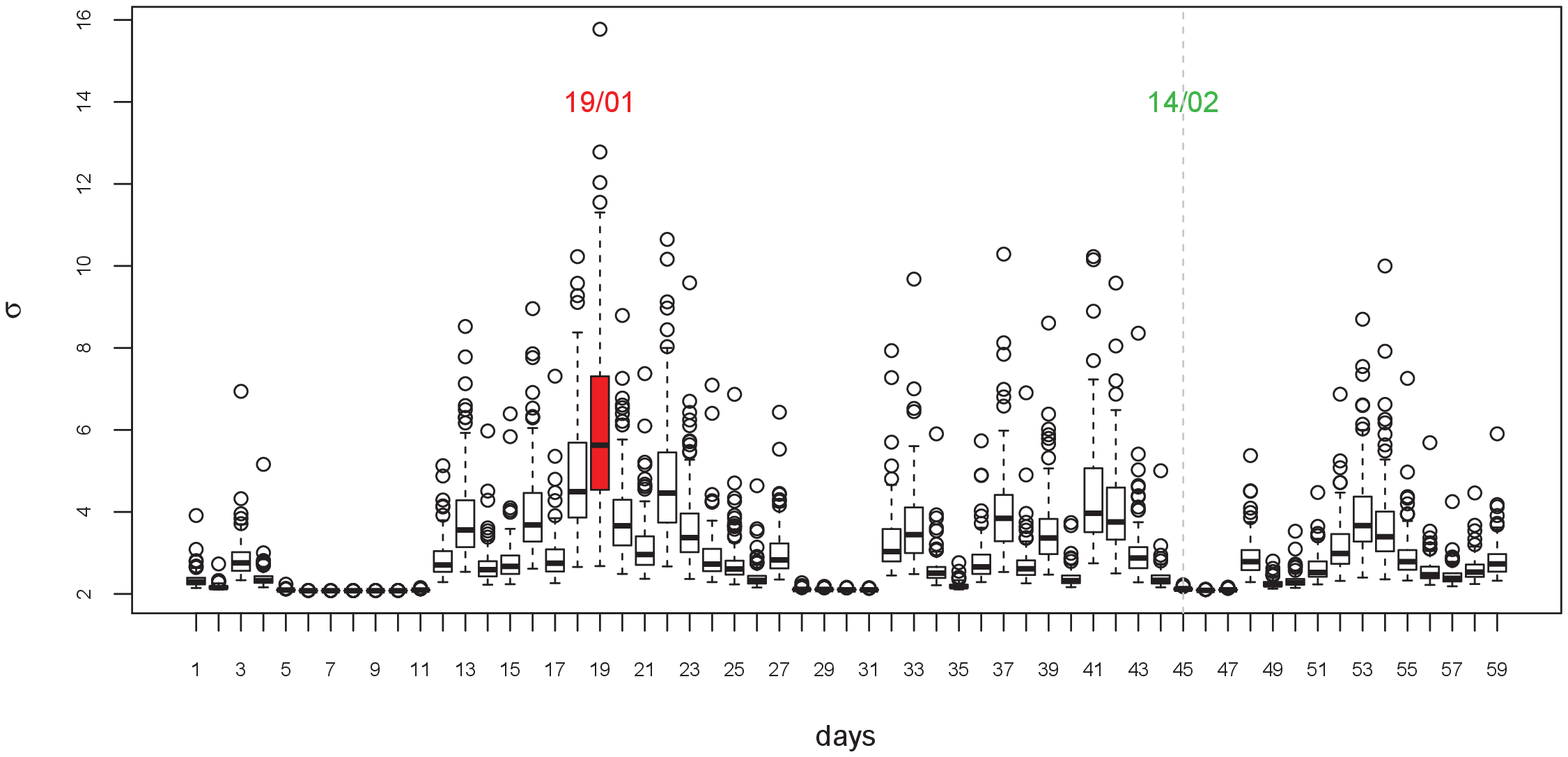}
\includegraphics[width=7cm]{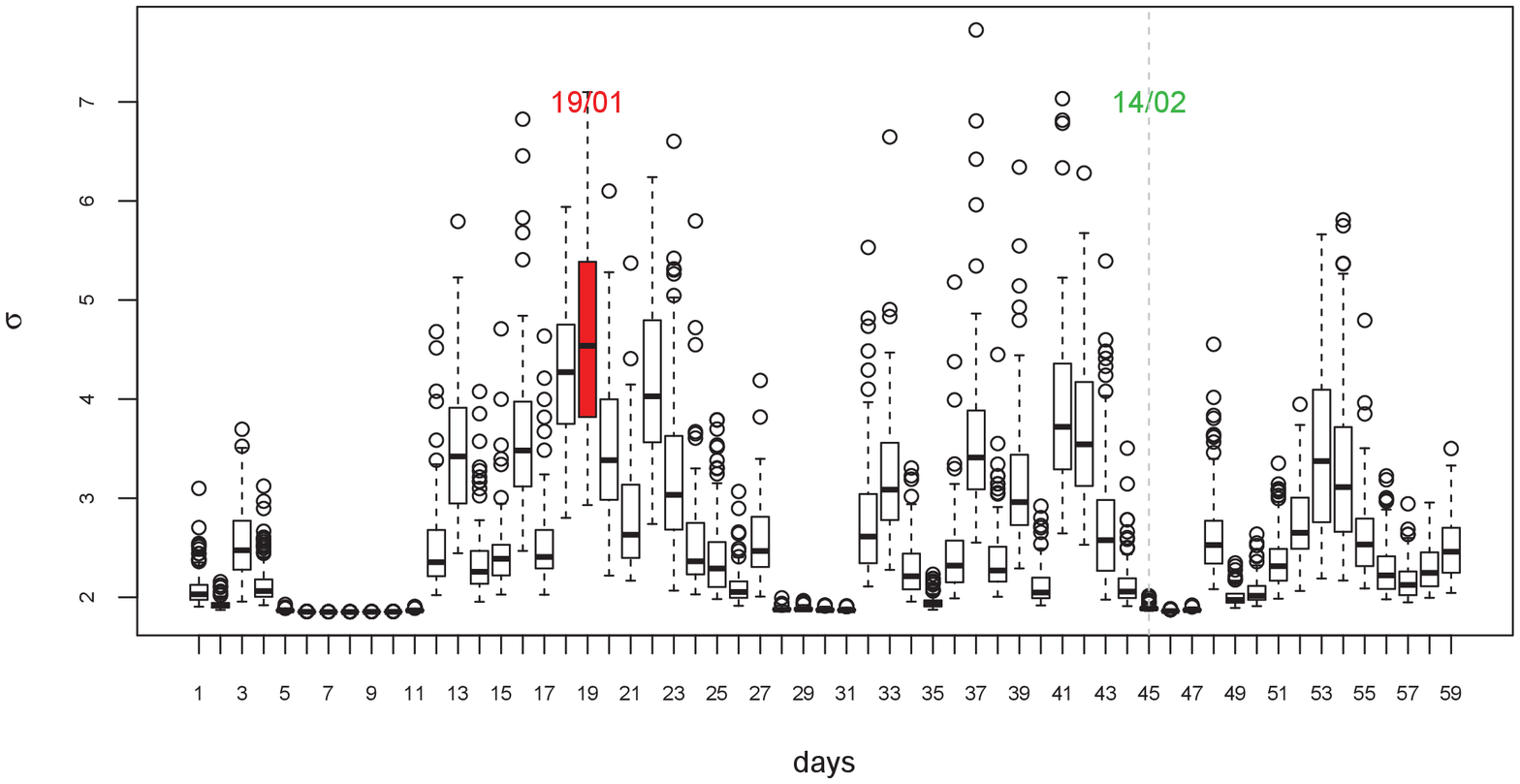}
\caption{\label{boxplot} Boxplot of the posterior means of $\sigma_y(i,j)$ (left) and $\sigma_x(i,j)$ (right) }
\end{figure}

 We observe that the posterior mean of $\kappa_y$ has a much smaller mean than the posterior mean of $\kappa_x$ which is consistent with the fact that, in general, simulated data are shifted to the right in relation to the observed data, indicating the possible existence of some bias in the simulated data. The posterior mean of the precision (inverse of the variance) parameters for the space model ($\tau_W$) and for the temporal model ($\tau_Z$) suggest that time dependence is stronger than space dependence. The posterior mean for $\beta_y$ is slightly smaller than the posterior mean for $\beta_x$. This naturally contributes for higher values for $\sigma_y(i,j)$ relatively to $\sigma_x(i,j)$ and with greater dispersion, as it can be seen in  Figure  \ref{boxplot} where we show daily boxplots  of the posterior means of the parameters $\sigma(i,j),\forall j$  for both models. In that figure it is marked two dates, 19 of January, a day where it was observed a storm with heavy winds (storm GONG, maximum observed wind 29.6m/s), particularly in regions close to the littoral, and 14th of February, a very mild day all over the country (Valentine's day; maximum observed wind 8.20m/s).  The variation observed along the days is consistent with the fact that on windy days the  maximum wind speed  along the stations varies much more than on mild days. Also the temporal dependence is clear in these pictures.

\begin{table}[ht]
\centering
\caption{Summary statistics  for the marginal posterior distributions} \label{tab1}
\begin{tabular}{cccccccc}
  \hline
 & mean & standard deviation & 2.5\% quantile & median & 97.5\% quantile & min & max \\
  \hline
$\alpha$ & 0.451 & 0.042 & 0.346 & 0.462 & 0.499 & 0.241 & 0.500 \\
 $\beta_y$ & -1.094 & 0.149 & -1.376 & -1.093 & -0.806 & -1.541 & -0.595 \\
  $\beta_x$ & -0.854 & 0.134 & -1.105 & -0.849 & -0.598 & -1.243 & -0.365 \\
  $\kappa_y$ & 5.312 & 0.197 & 4.936 & 5.310 & 5.701 & 4.675 & 5.936 \\
  $\kappa_x$ & 18.588 & 0.725 & 17.230 & 18.585 & 20.080 & 16.480 & 20.930 \\
  $\tau_W$ & 4.234 & 0.715 & 2.923 & 4.187 & 5.762 & 2.218 & 7.022 \\
  $\tau_Z$ & 0.396 & 0.095 & 0.241 & 0.384 & 0.618 & 0.191 & 0.850 \\
  $\xi_y$ & -0.070 & 0.002 & -0.074 & -0.070 & -0.067 & -0.077 & -0.065 \\
  $\xi_x$ & -0.081 & 0.001 & -0.084 & -0.081 & -0.078 & -0.085 & -0.076 \\
   \hline
\end{tabular}
\end{table}

\begin{figure}[!ht]\centering
\includegraphics[width=5.3cm]{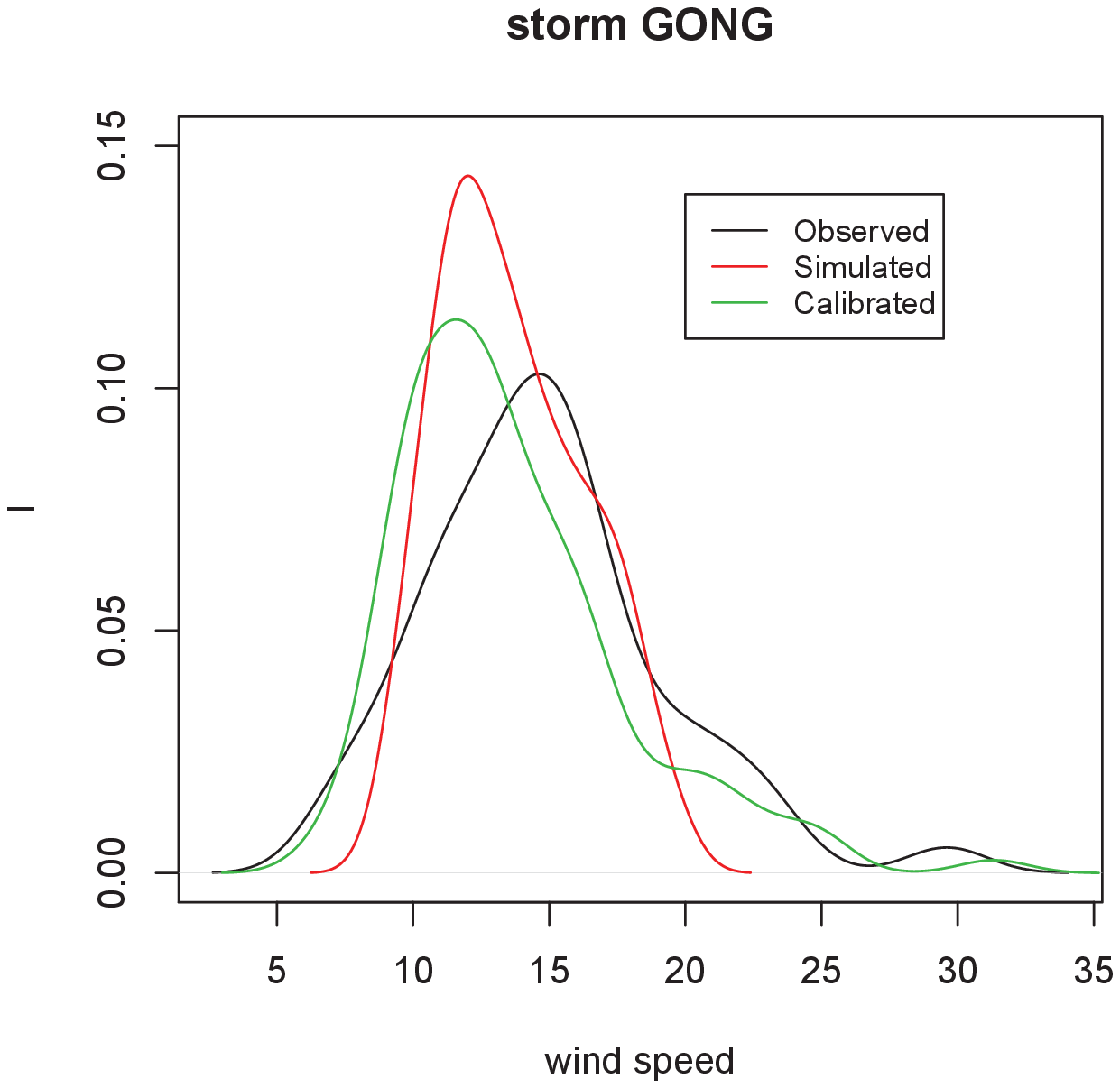}
\includegraphics[width=5.3cm]{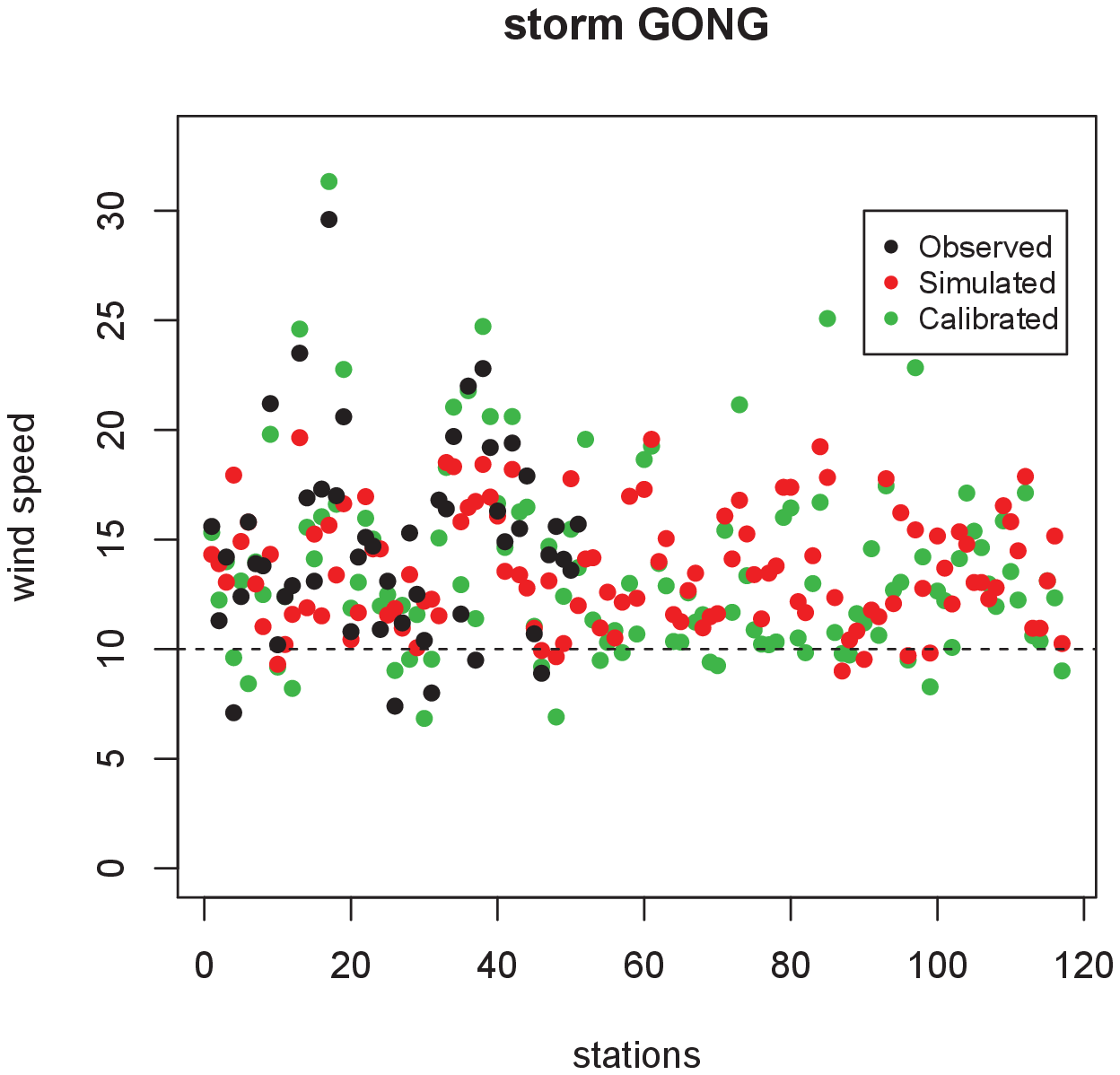}
\caption{\label{gong}   Kernel density estimation (left), observed and simulated maximum wind speed for each station, together with the mean of the predictive distribution for the calibrated data, for a storm day.}
\end{figure}

\begin{figure}[!ht]\centering
\includegraphics[width=5.3cm]{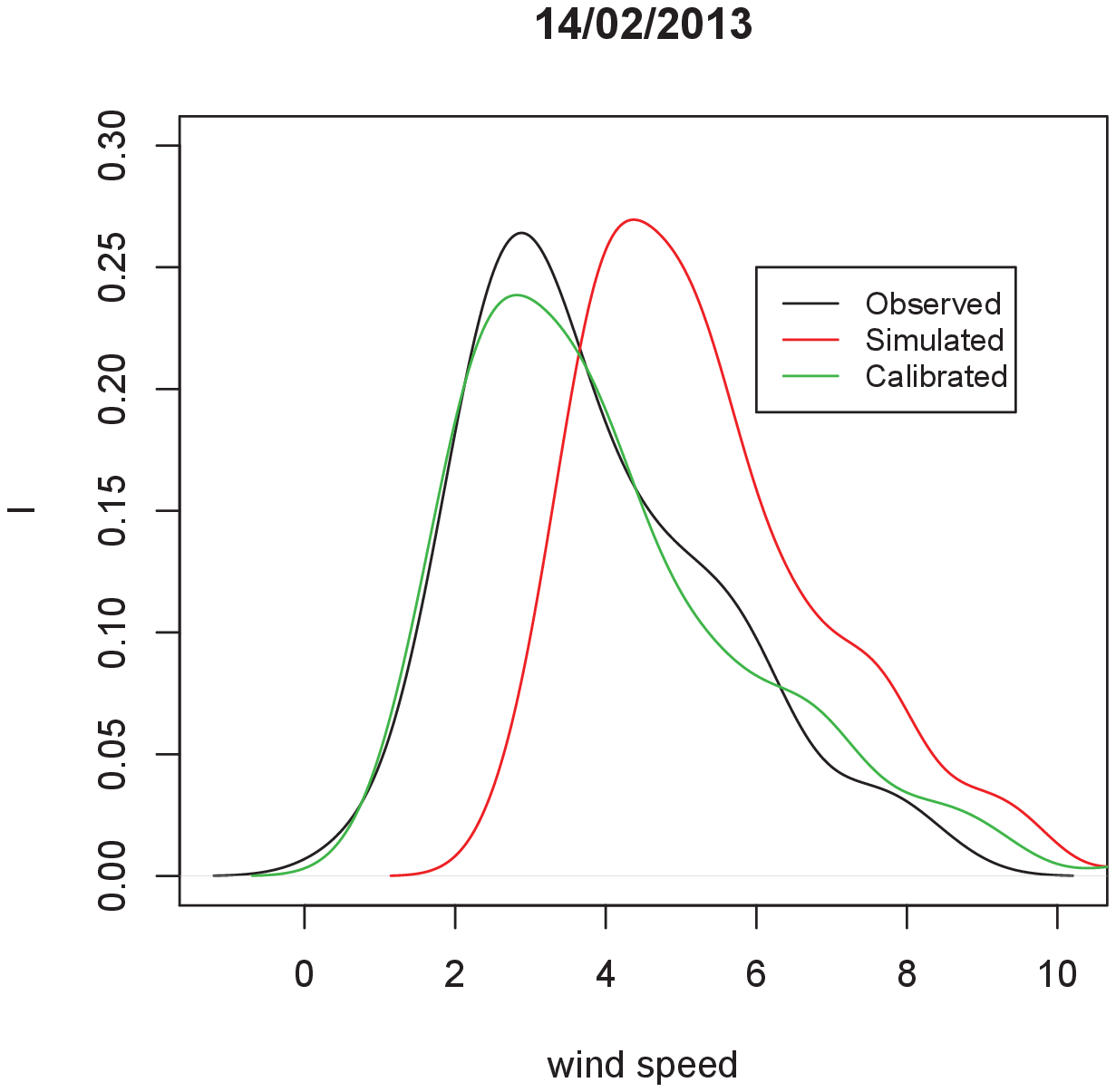}
\includegraphics[width=5.3cm]{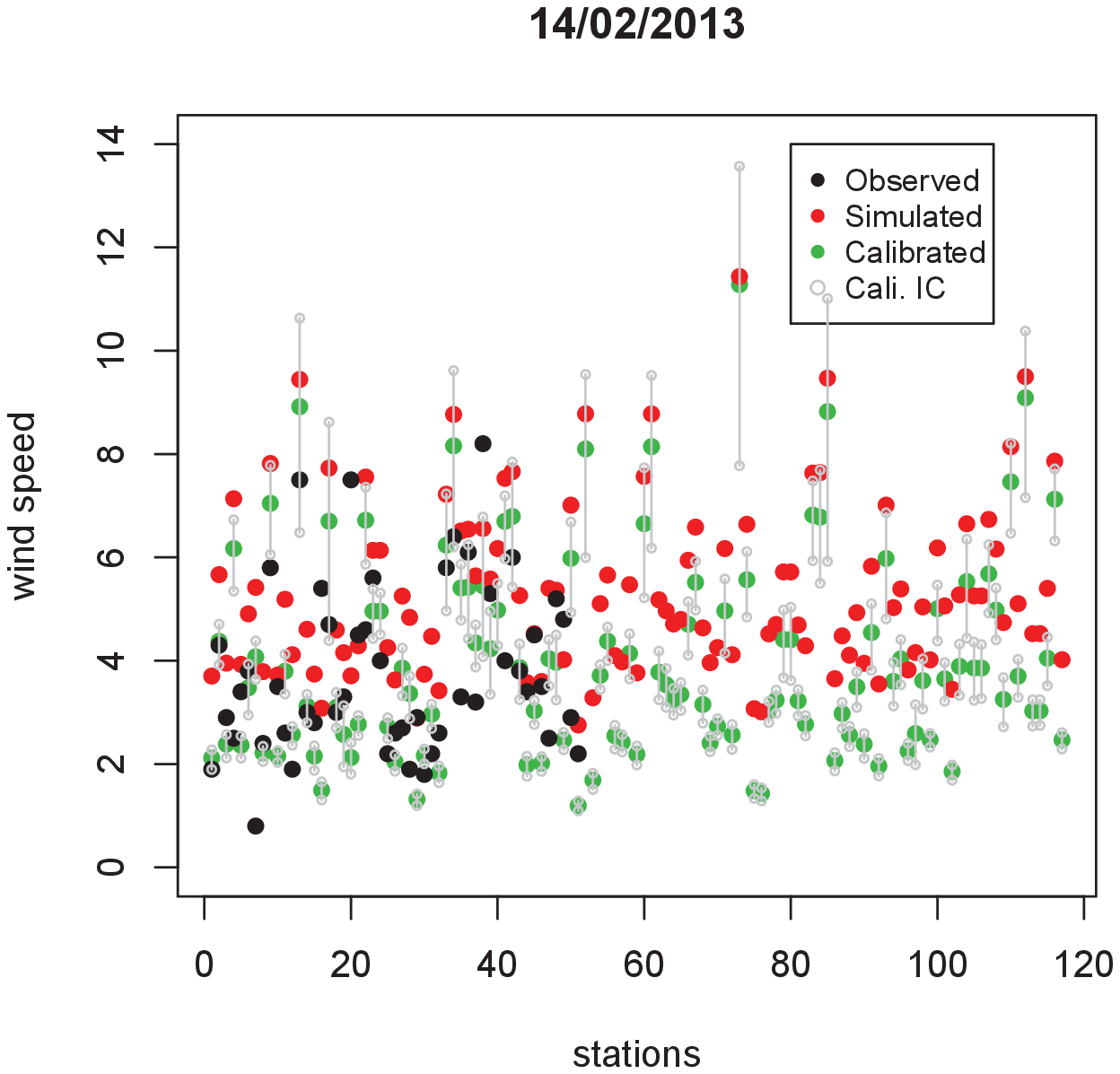}
\caption{\label{val}  Kernel density estimation (left), observed and simulated maximum wind speed for each station, together with the mean of the predictive distribution for the calibrated data, for a mild day.}
\end{figure}

These two days were studied, in particular, for exemplification of the conditional quantile calibration method proposed. For the purpose of exemplification of the results we represent in Figures \ref{gong} and \ref{val}, on the left, a kernel density estimation (considering all the stations) for the observed and simulated maximum wind speed on that day, together with  the mean of the predictive distribution of the calibrated data as defined in (\ref{eqmb}). On the right side we represent the observed and simulated maximum wind speed on that day for each station, together with the mean of the predictive distribution for the calibrated data.

We observe that, on a storm day (Figure \ref{gong}) the observed winds have longer tails than simulated winds. The calibration method was able to capture both tails of the distribution for the observed data, although it shifted the bulk of the distribution to the left.  Regarding a mild windy day (Figure \ref{val}),  the distribution of the simulated data is shifted to the right relatively to the distribution of the observed data with longer tails, as it was observed in a preliminary study. This bias is corrected with the calibration method.

\begin{figure}[!ht]\centering
\includegraphics[width=10cm]{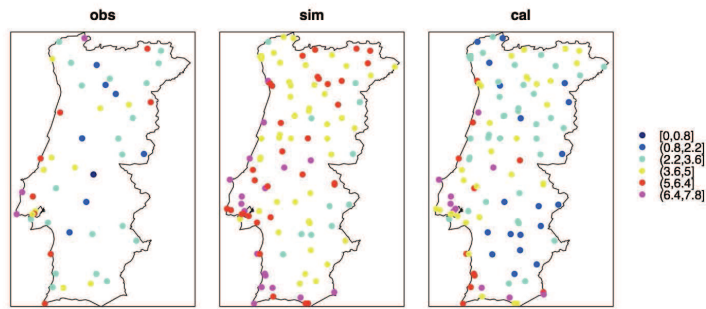}
\caption{\label{mapa_gong} Storm day: observed, simulated and calibrated maximum wind speeds}
\end{figure}

\begin{figure}[!ht]\centering
\includegraphics[width=10cm]{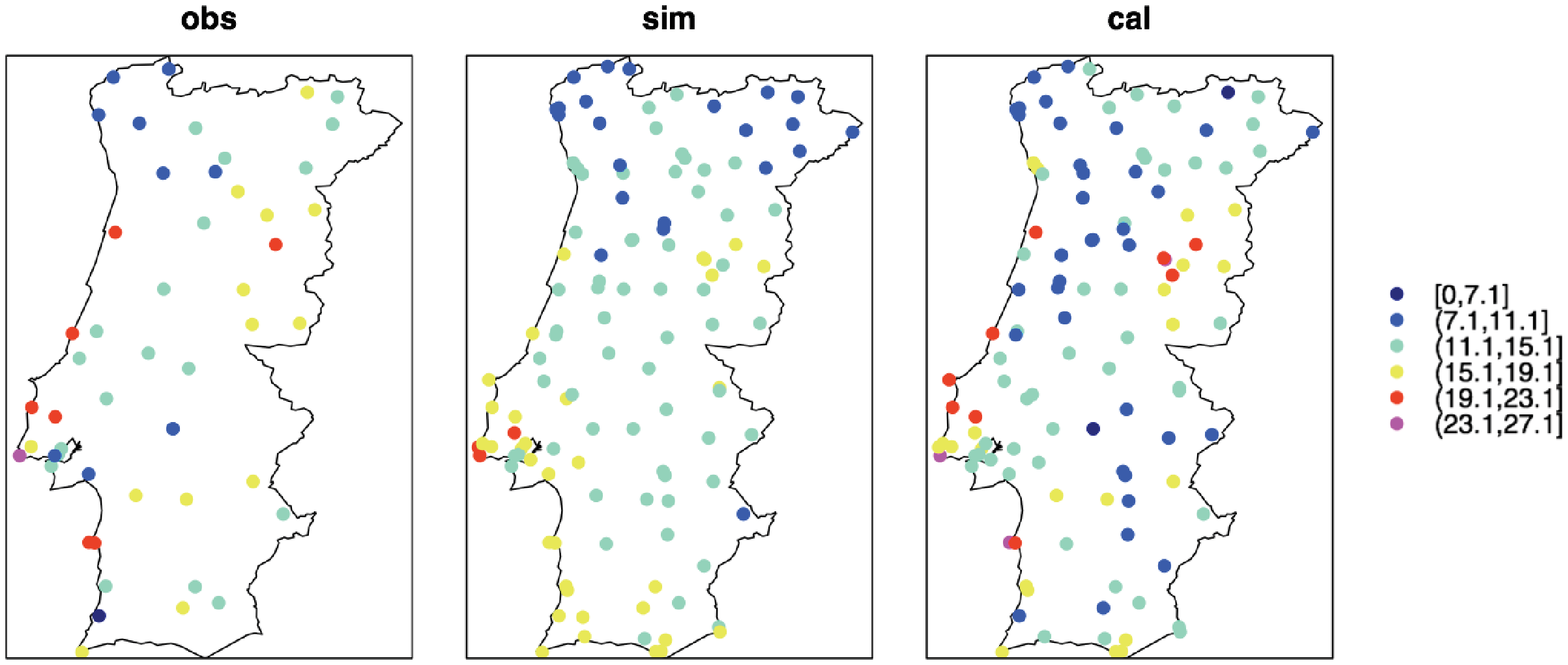}
\caption{\label{mapa_val} Mild day: observed, simulated and calibrated maximum wind speeds }
\end{figure}

In Figures \ref{mapa_gong} and \ref{mapa_val} there is a spatial representation of the observed, simulated and calibrated values for each of these two days.

\section{Discussion and further extensions}\label{disc}

In this article we discussed several possible ways of calibrating data obtained from a simulator based on observations at stations. We also  proposed and implemented a conditional quantile matching calibration (CQCM) using a space-time extended generalized Pareto distribution.

The  performance of the CQCM method was exemplified with two specific days, a storm day and a mild day. In both cases the calibrated data matched well the observed data on the tails, although on the storm day it did not capture well the bulk of the distribution. Ideally this method should be extended to the grid level, since the simulator produces data at a fine grid level and this is much more interesting if the objective is the construction of a risk map. However this extension is not trivial and some assumptions regarding the model structure have to be assumed.

 Damages in electricity grid are basically governed by extreme winds and primarily simulated and observed data coming from the right tail differ.
     Hence adequate calibration methods must be specifically adapted to extreme observations coming from the right tails and
     methods and models to be used in calibration should ideally be compatible with extreme value theory.
    A range of approaches for characterising the extremal behaviour of spatial process  have been suggested and a brief comparison of these methods can be found in Tawn \emph{et al.}~(2018).
         Downscaling method described by Towe {\it et al.}~(2017)--- based on the  conditional extremes process---is more suitable, with adequate modifications, to calibrate extreme simulated data based on observed wind speeds. Work on this approach is under progress.
\section*{Acknowledgments}
Research partially financed by national funds through FCT -  Funda\c{c}\~{a}o para a Ci\^{e}ncia e a Tecnologia, Portugal, under the projects  PTDC/MAT-STA/28649/2017 and UIDB/00006/2020.

\References

\item[] Aitchison, J. and Dunsmore, I.~R. (1975).  \emph{Statistical Prediction Analysis}. Cambridge: Cambridge University Press.
\item[] Beirland, J., Goegebeur, Y., Segers, J. and Teugels, J. (2004). \emph{Statistics of
Extremes: Theory and applications}. J Wiley, Chichester.
\item[] Banerjee, S., Carlin, B.~P. and Gelfand, A.~E. (2004). \emph{Hierarchical Modeling and Analysis for Spatial Data}, Boca Raton, FL: Chapman and Hall.
\item[] Berrocal, V.~J. (2019). Data assimilation. In Gelfand AE, Fuentes M, Hoeting JA and Smith RL, \emph{Handbook of Environmental and Ecological Statistics}, 133\,--\,151, Chapman and Hall/CRC.
\item[] Berrocal, V.~J., Gelfand, A.~E. and Holland, D.~M. (2012).  Space-time data fusion under error in computer model output: An application to modeling air quality. \emph{Biometrics}, {\bf 68},
      837\,--\,848.
\item[] Berrocal, V.~J., Gelfand, A.~E. and Holland, D.~M. (2014), Assessing Exceedance of Ozone Standards: A Space Time Downscaler for Fourth Highest Ozone Concentrations,
\emph{Environmetrics}, {\bf 25}, 279\,--\,291.
\item[] Cardoso, R.~M., Soares, P.~M.~M, Miranda,P.~M.~A. and Belo-Perira, M. (2013), WRF High Resolution Simulation of Iberian Mean and Extreme Precipitation Climate, \emph{International
Journal of Climatology}, {\bf33}, 2591\,--\,2608.
\item[] De Zea Bermudez, P., and S. Kotz. 2010. Parameter estimation of the generalized Pareto distribution—Part I. \emph{J. Stat. Plan. Inference}, {\bf 140(6)}, 1353\,–-\,1373.
\item[] Foley,K.M. and Fuentes, M.  (2008). A Statistical Framework to Combine Multivariate Spatial Data and Physical Models
for Hurricane Surface Wind Prediction. \emph{Journal of Agricultural, Biological, and Environmental Statistics}, {\bf 13},37\,--\,59.
\item[] Fuentes, M. and Raftery, A.~E. (2005).  Model evaluation and spatial interpolation by Bayesian
combination of observations with outputs from numerical models. \emph{Biometrics}, {\bf 61},
      36\,--\,45.
\item[] Kennedy, M. and O'Hagan, A. (2001). Bayesian Calibration of Computer Models.  {\it Journal of the Royal Statistical Society, Series B}, {\bf 63},
      425\,--\,464.
\item[] Kalnay (2003) \emph{Atmospheric modeling, data assimilation and predictability}. Cambridge University Press.
\item[] Kang, P., Koo, C. and  Roh, H. (2017). Reversed inverse regression for the univariate linear calibration
and its statistical properties derived using a new methodology. \emph{International Journal of
Metrology and Quality Engineering}, {\bf 8 (28)}.
\item[] Lavagnini, I. and Magno, F.(2007). A statistical overview on univariate calibration, inverse regression, and detection limits: application to gas chromatography/mass  spectrometry technique. \emph{Mass Spectrometry  Techniques}, {\bf 26}, 1\,--\,18
\item[] McMillan N.~J., Holland, D.~M., Morara, M., and Feng J. (2010). Combining numerical model output and particulate data using Bayesian space-time modeling. \emph{Environmetrics},{\bf 21}, 48\,--\,65.
\item[]  Michelangeli, P.~A., Vrac, M., and Loukos, H. (2009).
Probabilistic downscaling approaches: Application to wind
cumulative distribution functions. \emph{Geophys. Res. Lett.}, {\bf{36}},
1\,--\,6.
\item[] Muehleisen, R.~T. and Bergerson,  J. (2016). Bayesian Calibration - What, Why And How. \emph{International High Performance
Buildings Conference}. Paper 167.
\item[] Naveau P., Huser R., Ribereau P. and Hannart A. (2016). Modeling jointly low, moderate, and heavy rainfall intensities
without a threshold selection. \emph{Water Resources Research}. 2753\,--\,2769.
 \item[] Osborne, C. (1991). Statistical Calibration: A Review. \emph{International Statistical Review / Revue Internationale De Statistique}, {\bf 59}, 309\,--\,336.
\item[] Pereira, S., Pereira, P., de Carvalho, M. and de Zea Bermudez, P. (2019). Calibration of extreme values of simulated and real data. \textit{Proceedings of International Workshop on Statistical Modelling 2019}.
\item[] Rue, H. and Held, L. (2005). \emph{Gaussian Markov Random Fields: Theory and Applications}. Monographs on Statistics and Applied Probability, vol. 104. Ghapman\& Hall: London.
\item[]  Racine-Poon, A. (1988). A Bayesian Approach to non-linear calibration problems. \emph{JASA}, \textbf{83}, 650\,--\,656.
\item[] Skamarock, W.C~., Klemp, J.B~., Dudhia, J., Gill, D.~O., Barker, D.~M., Duda,
M.~G., Huang, X.~Y., Wang, W. and Powers, J.~G. (2008). A description of the
advanced research WRF version 3 . \emph{NCAR Tech. Note TN-475\_STR}.
\item[] Sigrist, F.,  K\"{u}nsch, H.R. and Stahel, W.A. (2015). Stochastic partial differential equation based modelling of large space-time data sets. \emph{Journal of the Royal Statistical Society, Series B}, {\bf 77},
      3\,--\,33.
\item[] Sturtz, S.,  Ligges, U. and  Gelman, A. (2005). R2WinBUGS: A Package for Running WinBUGS from R. \emph{Journal of Statistical Software},
{\bf 12}, 1\,--\,16.
\item[] Towe, R.P., Sherlock, E.F., Tawn, J.A., Jonathan, P. (2017). Statistical downscaling for future extreme wave heights in the North Sea.
  \emph{Annals of Applied Statistics}. {\bf 11}, 2375\,--\,2403.
\item[] Wilkison, R.~D. (2010).  Bayesian Calibration of expensive multivariate computer experiments. \emph{In Computational Methods for Large scale inverse problems and quantification of uncertainity}. J. Wiley and Sons.
\item[] Zidek, J.V., Le, N.~~D. and Liu, Z.  (2012). Combining data and simulated data for space-time fields: Application to ozone. \emph{Environmental and Ecological Statistics}, {\bf 19}, 37\,--\,56.
\endrefs

\end{document}